\documentclass[10p]{iopart}
\usepackage{iopams}
\expandafter\let\csname equation*\endcsname\relax
\expandafter\let\csname endequation*\endcsname\relax

\usepackage{amsmath,empheq}
\usepackage{amsfonts}
\usepackage{amsthm}
\usepackage{amssymb}
\usepackage{graphicx}
\usepackage{hyperref}
\usepackage{soul}
\usepackage{harvard}
\usepackage[normalem]{ulem}
\usepackage{textcomp}
\usepackage{bbm}
\usepackage{cite}
\usepackage{commath} 
\usepackage{xcolor}
\usepackage{mathtools}
\usepackage[symbol]{footmisc}
\usepackage{tikz}
\usepackage{tikz-cd}
\usetikzlibrary{%
  matrix,%
  calc,%
  arrows%
}
\usepackage{pdflscape}
\usepackage{geometry}
	\newcommand{\ncd}{\newcommand}
	\ncd{\mrm}    {\mathrm}
	\ncd{\beq} {\begin{equation}}
	\ncd{\eeq} {\end{equation}}
	\ncd{\nn}{\nonumber}

	\def\d{{\rm d}}

	\def\basis[#1]{\frac{\partial}{\partial #1}}
	\def\dt[#1]{\frac{\d}{\d #1}}

	\def\T{\mathcal{T}}

	\def\iiota{\dot\iota}

\begin{document}

\title[Contact polarizations  and associated metrics]{Contact polarizations and associated metrics in  geometric thermodynamics}	

	\author{C. S. Lopez-Monsalvo$^{1}$, F. Nettel$^{2,\sharp}$, V. Pineda-Reyes$^3$ \\and L. F. Escamilla-Herrera$^3$}
		\address{$^1$Conacyt-Universidad Aut\'onoma Metropolitana Azcapotzalco
		Avenida San Pablo Xalpa 180, Azcapotzalco, Reynosa Tamaulipas, 02200 Ciudad de 
		M\'exico, M\'exico}
		\address{$^2$Departamento de F\'isica, Fac. de Ciencias, Universidad Nacional Aut\'onoma de M\'exico, A.P. 50-542, 04510, CDMX, M\'exico.}
		\address{$^3$Instituto de Ciencias Nucleares, Universidad Nacional Aut\'onoma de M\'exico,
		A.P. 70-543, 04510, CDMX, M\'exico.}
		\eads{\mailto{$^\sharp$fnettel@ciencias.unam.mx}}

	\begin{abstract}
	
%In the geometric description of thermodynamics it is known that equilibrium states can be described by points on a contact manifold subject to a geometric constraint that can be identified with the first law of thermodynamics. This contact manifold is called the thermodynamic phase space and a curve joining two points on it constitutes a thermodynamic process. The first law is Legendre invariant, therefore it is natural to ask for the same kind of symmetries for the measurement of the length of such curves on the contact manifold. 

In this work we show that a  Legendre transformation is nothing but a mere  change of contact polarization from the point of view of contact geometry. Then, we construct a set of Riemannian and pseudo-Riemannian metrics on a contact manifold by  introducing  almost contact and para-contact structures and  we analyze their isometries. We show that it is not possible to find a class of metric tensors which fulfills two properties: on the one hand, to be polarization independent i.e. the Legendre transformations are the corresponding  isometries and, on the other,  that it induces a Hessian metric into the corresponding  Legendre submanifolds. This second property is motivated by the well known Riemannian structures of the geometric description of thermodynamics which are based on Hessian metrics on the space of equilibrium states and whose properties are related to the fluctuations of the system. We find that to define a Riemannian structure with such properties it is necessary to abandon the idea of an associated metric to an almost contact or para-contact structure. We find that even extending the contact metric structure of the thermodynamic phase space the thermodynamic desiderata cannot be fulfilled.  
  
	\end{abstract}

\maketitle

%%-----------------------------------------------------------

\section{Introduction}

Contact geometry is the natural setting to describe processes satisfying a constraint akin to the first law of thermodynamics.  A  thermodynamic process is realized as a smooth curve connecting two points on a contact manifold. Each point of the manifold is referred as a thermodynamic state. In this sense, we call a contact manifold a thermodynamic phase space (TPS) where the constraint given by the first law defines its contact structure \cite{HamThermo,2018Schaft}. 

A contact structure on an odd-dimensional manifold is a maximally non-integrable distribution of co-dimension one  hyperplanes. That is, for every pair of points on the manifold, there exists a curve connecting them such that its tangent vector at each point along the path lies  in a hyperplane of the distribution. Furthermore, in every open set around each point on the manifold, there is a set of local coordinates such that the 1-form generating the contact distribution resembles the first law of thermodynamics [see \eqref{eq.eta1}, below]. This motivates  us to call  those curves whose tangent vector at each point is annihilated by  this 1-form, thermodynamic processes \cite{bravetti2015conformal,2018Schaft}.

From a thermodynamic perspective, maximal dimensional embedded integral submanifolds such that its tangent bundle is entirely contained in the contact distribution correspond to the realization of specific thermodynamic systems. That is, those where the embedding is defined by a  thermodynamic fundamental relation of a given set of independent thermodynamic variables, e.g. entropy, volume, temperature, pressure, volume, etc \cite{1985Callen,2018Schaft}. In the context of geometric thermodynamics these are called \textit{equilibrium spaces} \cite{MRUGALA91,MRUGALA1978,2013Arno}. Such submanifolds, when equipped with a metric whose components correspond to the Hessian of the fundamental relation,  are the subject of geometric thermodynamics \cite{shima2007,GarciaAraiza2018}. In particular, it has been established that the  Riemannian structure corresponding to a specific choice of thermodynamic potential encodes information stemming from  thermodynamic fluctuation theory \cite{Goto2015,Weinhold1975,2009Weinhold,1979Rupp,Ruppeiner1995}. 

Each thermodynamic potential defines a different Legendre submanifold and, since its metric description arises from fluctuation theory, it yields a different Riemannian structure, signalizing a non-equivalence of ensembles in the geometric characterization of fluctuations \cite{termometrica}. For instance, it may occur  that different potentials may yield non Hessian metrics \cite{1984Salamon}. Thus, albeit a thermodynamic system in equilibrium does not depend on the choice of thermodynamic potential describing it, its geometric structure -- related to its fluctuations -- does \cite{Santoro2005,2010Liu}. This motivated the search for a potential independent form of the Riemannian structures for the equilibrium spaces, that is, metrics sharing the Legendre symmetry of the thermodynamic description ~\cite{2007Quev,Pineda2019,termometrica,2014GarciaPelaez}. 

In this manuscript we show that, when viewed from the thermodynamic phase space,  each potential choice corresponds to a contact polarization for the contact distribution. Indeed, given a fundamental relation written in terms of  a particular thermodynamic potential, the corresponding embedding provides a specific choice of `position' and `momentum' coordinates for the symplectic structure of the contact planes. Thus, a Legendre transformation, interchanging the role of a pair of conjugate thermodynamic variables,  can be understood as a change of symplectic polarization \cite{2001ArnoldBook}.  As expected, this is a symmetry of the contact structure. We also obtain a class of metrics in the thermodynamic phase space invariant under a change of symplectic polarizations, generalizing the metric contact structure.  

This work is structured as follows: In section \ref{sec:Heiseberg}, we explore the thermodynamic phase space in terms of a basis of vector fields satisfying the Heisenberg algebra commutation relations and define the notion of contact polarization. In section \ref{sec:symmetries}, two different kind of horizontal contact Hamiltonians and the symmetry transformations generated by their corresponding contact Hamiltonian vector fields are explored, namely rotation and  polarization scalings. In section \ref{sec:ACS-APCS}, we introduce  the canonical almost contact structure and three different almost para-contact structures. We  study their symmetry properties under the transformations generated by the contact Hamiltonian vector fields of the horizontal contact Hamiltonians. In section \ref{sec:Metrics-ACS-APCS},   metrics for the contact manifold are constructed using the almost contact and almost para-contact structures and their properties are analyzed. In section \ref{sec:CRA} we present an automorphism on the tangent spaces of the contact manifold which allows us to construct a family of metrics for which the Legendre transformations are a set of isometries. Finally, in section \ref{sec:conclusions} we provide some closing remarks and discuss the main conclusions of this work.

\section{Thermodynamic Phase Space and contact polarizations}  \label{sec:Heiseberg}

Let us consider a $(2n+1)$ dimensional manifold $\T$ together with a set of vector fields providing a basis for each tangent space such that  they satisfy the commutation relations of the $n$th Heisenberg group algebra \cite{Ivanov2007}. That is, for each point $p\in\T$, the basis for the tangent space $T_p\T$ is given by the linearly independent  $2n+1$ vector fields  $\{Q_a,P^a,\xi\}$ whose Lie bracket satisfies  
	\beq
	\label{eq.HCM}
	[P^a,Q_b] = \delta^a_{\ b} \xi, \quad [\xi,P^a]=0 \quad \text{and} \quad [\xi,Q_a] = 0.
	\eeq  
This is the simplest example of a bracket generating distribution \cite{kobayashi1963}, namely, let $\mathcal{D}\subset T\T$ be the $2n$-dimensional distribution generated by $\{P^a,Q_a\}_{a=1}^n$, then
	\beq
	\label{eq.bracketgenerating}
	T\T = \text{span}\left(P^a,Q_a,[P^a,Q_b] \right).
	\eeq
Condition \eqref{eq.bracketgenerating} implies that any two points $p,q \in \T$ can be joined by a curve  such that at each point along the path its tangent vector lies in $\mathcal{D}$.

Historically, such condition was realized in connection with thermodynamic processes in the following sense: if each point in the manifold $\T$ corresponds to a possible thermodynamic state  characterized by $2n+1$ quantities, then for any two states $p,q\in \T$ there is a  process  joining them such that at each point along the path  the first law of thermodynamics is satisfied.  Indeed,  $\mathcal{D}$ is a contact distribution corresponding to the kernel of a 1-form $\eta\in T^*\T$. Thus, around each point $p\in\T$ there is a local set of coordinates $\{q^a,w,p_a\}$ in which the 1-form $\eta$ is written as \cite{blair2010}
	\beq
	\label{eq.eta1}
	\eta = \d w - \sum_{a=1}^n p_a \d q^a.
	\eeq 
This is known as Darboux theorem.

Any vector field $X\in T\T$ such that
	\begin{align}
	\eta(X) 	& = \eta\left[X^w \frac{\partial}{\partial w} + \sum_{a=1}^n \left(X^a_q \frac{\partial}{\partial q^a} + X^p_a \frac{\partial}{\partial p_a} \right)\right]\nonumber\\
			& = X^w -\sum_{a=1}^n p_a X^a_q\nonumber\\
			& = 0,
	\end{align}
is a combination of $2n$ vector fields, that is
\beq
X=\sum_{a=1}^n \left[X^a_q Q_a + X^p_a P^a \right],
\eeq
where
	\beq
	\label{eq.basis}
	P^a=\frac{\partial}{\partial p_a}, \quad \text{and}\quad  Q_a=\frac{\partial}{\partial q^a} + p_a \frac{\partial}{\partial w},
	\eeq
while $X^w$, $X_q^a$ and  $X^p_a$ are the corresponding components of $X$ in this basis.
It is straightforward to verify that the Lie bracket of the vector fields $\eqref{eq.basis}$ satisfies the Heisenberg commutation relations \eqref{eq.HCM}, that is
\beq\label{EQ.nonint}
\left[P^a,Q_b \right] =\delta^a_{\ b }\xi,
\eeq
while the vector field $\xi$ satisfies
	\beq
	\label{eq.reeb}
	\d \eta(\xi,X) = 0 \quad \text{and} \quad \eta(\xi) = 1 
	\eeq
for any vector field $X \in T\T$. In the literature, the vector field satisfying \eqref{eq.reeb} is called the Reeb vector field \cite{blair2010}.

The restriction of the exterior derivative of $\eta$ to $\mathcal{D}$   yields \cite{2001ArnoldBook}
	\beq \label{simplecticstructure}
	\left.\d \eta\right\vert_\mathcal{D}(X,Y) = -\sum_{a=1}^n \d p_a \wedge \d q^a (X,Y)  = \Omega(X,Y)
	\eeq
for any pair $X,Y \in \mathcal{D}_p$, where $\Omega$ is a bilinear anti-symmetric form for $\mathcal{D}|_p$, providing it with a symplectic structure. Thus, at each point $p\in\mathcal{T}$, the vector space $\mathcal{D}\vert_p$ is a symplectic space. A Legendre sub-manifold $\mathcal{E}$ of $(\mathcal{T,\eta})$ is defined by the conditions $T\mathcal{E} \subset T \mathcal{T}$, where $\eta\vert_\mathcal{E} = 0$ means that for any tangent vector $V\in T_p \mathcal{E}$, $\eta(V) = 0$, together with ${\rm dim}(\mathcal{E})=n$ . It is straightforward to verify that $T\mathcal{E}\subset \mathcal{D}$ is in involution. Such  conditions, however, do not specify a unique sub-manifold, but $2^n$ distinct embeddings $\ell:\mathcal{E}\longrightarrow\mathcal{T}$ (cf. Appendix 4 of \cite{2013Arno}). To see this, consider the following diagram

	\beq
	\label{diag.bundle}
	\begin{tikzpicture}[]
	\matrix[matrix of math nodes,column sep={45pt,between origins},row
	sep={50pt,between origins},nodes={asymmetrical rectangle}] (s)
	{
						&  		 	& |[name=a2]|  T\mathcal{E} \subset \mathcal{D} \supset	T\mathcal{E'} &  	 & |[name=a3]| 	\\[-35pt]
						&			& \cap					&		 &						\\[-35pt]
						& 			& |[name=b2]| T\mathcal{T}	& 		 &		 				\\
						& 			& |[name=c2]| \mathcal{T}		&		 & 						\\
	|[name=c1]| \mathcal{E}	&			&						&		 & |[name=c3]| \mathcal{E}'	\\
	}
	;
	\draw[->]
			(a2) edge[bend right= 30] node[left] {\(\pi_{\mathcal{E}}\)} (c1)
			(a2) edge[bend left=30] node[right] {\(\pi_{\mathcal{E}'}\)} (c3)
			(c1) edge node[auto] {\(\ell\)} (c2)
			(c3) edge node[above] {\(\ell'\)} (c2)
			(b2) edge node[auto] {\(\pi_{\mathcal{T}}\)} (c2)
	;
	\end{tikzpicture}
	\eeq
illustrating two such Legendre submanifolds. Here, $\pi_\mathcal{E}$, $\pi_{\mathcal{E'}}$ and $\pi_{\mathcal{T}}$ represent the  projections of the corresponding tangent bundles. Note that, Legendre sub-manifolds are not necessarily diffeomorphic to one another. Moreover, at every point $p\in\mathcal{T}$, we say $T_p\mathcal{E}$ is a \emph{Lagrangian} sub-space of $\mathcal{D}\vert_p$.  Recalling that a Lagrangian sub-bundle of the tangent bundle associated to a symplectic manifold is called a polarization if it is in involution, i.e. the set of sections of the Lagrangian sub-bundle is closed under the Lie bracket \cite{weinstein1977lectures}, in this manuscript we extend such definition to the case of contact manifolds as follows: a Legendrian sub-bundle $T\mathcal{E} \subset T\mathcal{T}$ is called a contact polarization if it is in involution. Therefore, a contact polarization is a foliation of the contact manifold $\mathcal{T}$ made with Legrende submanifolds  defined by the various thermodynamic potentials (cf. Appendix I).

\section{Horizontal Contact Hamiltonians and Symmetry Generators} \label{sec:symmetries}
As we have seen, the contact distribution $\mathcal{D}$ is generated by $\text{ker} \left( \eta\right)$. In fact, this distribution is given by an equivalence class of 1-forms $[\eta]$  with respect to the module of a conformal factor, ie. $\eta \sim \eta'$ if $\eta' = \lambda \eta$ with $\lambda$ a differentiable non-vanishing function on $\mathcal{T}$. Therefore, $\mathcal{D}$ is independent of the choice of $\eta$ in the same equivalence class. In fact, if a mapping $\Phi :\mathcal{T}\mapsto \mathcal{T}$ preserves $\mathcal{D}$, i.e., 
\beq
\Phi^* \eta=f_{\Phi}\eta, 
\eeq
we say that $\Phi$ corresponds to a contact transformation \cite{Boyer2011}. Here, $f_{\Phi}$ is a non vanishing function and where $\left[\Phi\right]^*: T^*\T \longrightarrow T^*\T$ is the \textit{pullback} induced map by $\Phi$. In the case $f_{\Phi}=1$, we say $\Phi$ is a strict contact transformation. Moreover, if the Lie derivative of the contact form along a vector field $X$  satisfies that
\beq \label{eq.ct}
\pounds _X \eta=\tau_X \eta, 
\eeq  
where $\tau_X:\mathcal{T}\mapsto \mathbb{R}$, we say that $X$ corresponds to an infinitesimal contact transformation. If, in addition,  
	\beq
	\pounds _X \eta =0,
	\eeq 
we say that the infinitesimal contact transformation is strict. Since the notion of a strict contact transformation depends on the chosen  1-form $\eta$ in the class generating the contact distribution, a strict contact transformation constitutes a symmetry of the contact form, alone.  

Consider a real valued function $h \in C^{\infty}(\T)$. A contact Hamiltonian system $(\mathcal{T},\eta, h)$ is defined by the relation \cite{2001ArnoldBook, HamThermo, Boyer2011}
\beq \label{eq.defham}
\eta(X_h) = h,
\eeq
where $X_h$ is a unique vector field called the contact Hamiltonian vector field. As $X_h$ generates a contact transformation it must satisfy equation \eqref{eq.ct}, then it can be shown that
\beq
\iiota_\xi \pounds_{X_h} \eta =\iiota_\xi \d h = \xi(h),
\eeq
where we use the notation $\iiota$ to denote the contraction operation between vector fields and differential forms. Thus we see that the diffeomorphism generated by $X_h$ is indeed a contact transformation
\beq \label{eq.lieeta}
\pounds_{X_h} \eta = \xi(h) \eta.
\eeq
Therefore, for $X_h$ to be a symmetry of the contact form $\eta$, it is sufficient that $\xi(h)$ vanishes, in such case we say that the contact Hamiltonian is a purely horizontal function.

Note that in our convention Hamilton's equations are expressed as
\beq
	\dot{q}^a=-\frac{\partial h}{\partial p_a}, \qquad \dot{p}_a=\frac{\partial h}{\partial q_a}+p_a\frac{\partial h}{\partial w}, \qquad a=1,\dots,n
\eeq
and
\beq
	\dot{w}=h-\sum_{b=1}^np_b\frac{\partial h}{\partial p_b}.
\eeq

\subsection{Legendre symmetry}

Let us consider  the contact Hamiltonian
	\beq
	\label{eq.conth}
	h_L = \frac{1}{2} \sum_{i=1}^m \left({q^i}^2 + {p_i}^2 \right).
	\eeq
Its corresponding contact Hamiltonian vector field is
	\beq
	X_{h_L} =  \sum_{i=1}^m \left[ \frac{1}{2} \left({q^i}^2 + {p_i}^2 \right) \xi + q^i P^i - p_i Q_i\right],
	\eeq
where we will use $i$, $j$, $k$, etc. to distinguish the indices that take values on a subset of the coordinates, $i,\, j,\, k = 1, \ldots, m$, where $m < n$, from those that take values on the complete set of coordinates, $a,\, b,\, c = 1,\ldots, n$. For the indices of the remaining coordinates we will use capital letters, $I,J = m +1, \ldots, n$.
Note that, restricted to $\mathcal{D}$, this is the generator of rotations in each contact plane, whose \emph{canonical} basis is given by  $\{Q_a,P^a\}_{a=1}^n$. Indeed, its flow generates the 1-parameter family of diffeomorphisms $\Phi_t:\T \longrightarrow\T$
	\beq
	{\Phi^m_t}= \left\{\begin{array}{ll}
				 w(t) 	 & = w_0 -\frac{1}{2} \sin(t)\sum_{i=1}^m \left[\left({{p_i}_0}^2 - {{q^i}_0}^2\right)\cos(t) + 2 \sin(t) {q^i}_0{p_i}_0 \right] \\
				 q^i(t) & = {q^i}_0\cos(t) - {p_i}_0 \sin(t)\\
				 p_i(t) & ={q^i}_0 \sin(t) + {p_i}_0 \cos(t),
				\end{array} \right.,
	\eeq
where ${q^i}_0=q^i(0)$ and ${p_i}_0=p_i(0)$; whilst mapping the rest of the coordinate functions into themselves.  In particular, we have that a  $\pi/2$-rotation in each contact plane of $\mathcal{D}$
	\beq  \label{legendre.transf}
	\Phi_{\frac{\pi}{2}}^m = \left\{\begin{array}{ll}
					w_(\frac{\pi}{2}) 	 & = w_0 - \sum_{i=1}^m {q^i}_0{p_i}_0 \\
					q^i(\frac{\pi}{2}) & =  -{p_i}_0 \\
					p_i(\frac{\pi}{2}) & ={q^i}_0 
					\end{array} \right.,
	\eeq
corresponds to a partial change of contact polarization and generates a partial Legendre transformation in $\T$. 

Since the contact Hamiltonian \eqref{eq.conth} is a purely horizontal function, it follows that the contact 1-form $\eta$ is propagated along the flow of $X_{h_L}$, that is 
	\beq
	\pounds_{X_{h_L}} \eta = 0.
	 \eeq
In particular, we have that
	 \beq
	 \left[{\Phi^m_{\frac{\pi}{2}}}\right]^*(\eta) = \eta,
	\eeq  
That is, the \emph{finite} transformation is a symmetry of the contact structure. Moreover, this is a Legendre involution (cf. subsection K of Appendix 4 in \cite{2013Arno}), that is, it carries a Legendre submanifold to a Legendre submanifold. This is obvious in the sense that the contact 1-form has $X_{h_L}$ as an infinitesimal symmetry. However, as we will shortly see, it is the discrete symmetries the ones which are relevant in the context of thermodynamics and, while infinitesimal symmetries imply the discrete case, the converse is not necessarily true. 

It is also useful to have the expressions for the Lie derivatives of the basis vectors. The case of the Reeb vector field is straightforward	
	\beq
	\pounds_{X_{h_L}} \xi = [X_{h_L},\xi]= -[\xi, X_{h_L}] = 0. 
	\eeq
		
Then, for $i,\, j = 1,\ldots, m$  we have   
\beq
\label{eq.lielq}
\pounds_{X_{h_L}} Q_i 	 = -P^i,
\eeq
while for $I = m +1, \ldots, n$ it is trivial that
\beq \label{eq.lielqalt2}
\pounds_{X_{h_L}} Q^I = 0.
\eeq

Similar calculations yield
\beq
	\label{eq.lielp}
	\pounds_{X_{h_L}} P^i = Q_i,
	\eeq
and
\beq \label{eq.lielpalt2}
\pounds_{X_{h_L}} P^I = 0,
\eeq
where we have used the canonical commutation relations \eqref{eq.HCM} to obtain the results. 

Now, let us consider the bundle isomorphism \cite{deLeon2019}
	\beq
	\flat(X) \equiv  \eta(X) \eta + \dot\iota_X \d \eta \quad \text{for} \quad X\in T\mathcal{T}.
	\eeq
The 1-forms dual to the horizontal vector fields $P^i$ and $Q_i$ correspond to $-\d q_i$ and $\d p^i$, respectively. Thus, the corresponding Lie derivatives are
	\begin{align}
	\label{eq.lieldq}
	\pounds_{X_{h_L}} \d q^i 	& = \iiota_{X_{h_L}} \d \left[\d q^i \right] + \d \left[\iiota_{X_{h_L}} \d q^i \right]\nonumber\\
						& = \d \left[\d q^i \left(\sum_{j=1}^m \frac{1}{2} \left({p_j}^2 + {q^j}^2 \right) \xi + q^j P^j - p_j Q_j \right) \right]\nonumber\\
						%& = \d\left[-p_i \right]\nonumber\\
						& = -\d p_i,
	\end{align}
and
	\beq
	\label{eq.lieldp}
	\pounds_{X_{h_L}} \d p_i = \d q^i,
	\eeq
while
	\beq \label{eq.lieldpa}
	\pounds_{X_{h_L}} \d q^I = 0,
	\eeq
and
	\beq \label{eq.lieldqa}
	\pounds_{X_{h_L}} \d p_I = 0,
	\eeq
for $I = m +1,\ldots, n$.
	
\subsection{Polarization scalings}

Consider now the contact Hamiltonian
	\beq
	h_S =  \sum_{a=1}^n q^a p_a.
	\eeq
In this case the contact Hamiltonian vector field is
	\beq  \label{eq.polscaling.X}
	X_{h_S} = \sum_{a=1}^n \left(q^a p_a\right)  \xi +\sum_{a=1}^n \left[ p_a P^a - q^a Q_a \right].
	\eeq
Again, this is clearly a contact symmetry [cf. equation \eqref{eq.lieeta}]
	\beq
	\pounds_{X_{h_S}} \eta = 0,
	\eeq
and its flow generates the 1-parameter family of anisotropic scalings $\delta_t:\T \longrightarrow \T$ 
	\beq
	\delta_t =\left\{\begin{array}{ll}
					w(t)   & = w_0\\
					q^a(t) & =  {q^a}_0 e^{-t} \\
					p_a(t) & = {p_a}_0 e^t 
					\end{array} \right..
	\eeq
In this case, the transformation acts on every coordinate of $\mathcal{D}$.
In particular we have
	\beq
	 \delta_{t}^*(\eta) = \eta,
	\eeq  
Note that, for $t>0$ one of the polarizations expands while the other shrinks. Thus, we call each member of the 1-parameter family $\delta_t$ a \emph{polarization scaling}.

Again, the action of the Lie derivative of the horizontal basis with respect to the generator $X_{h_S}$ is
%	\begin{align}
%	\pounds_{X_{h_S}} Q_a	& = \left[ \sum_{b=1}^n \left(q^b p_b\right)  \xi +\sum_{b=1}^n \left( p_b P^b - q^b Q_b\right),Q_a \right]\nonumber\\
%						& = -\left[Q_a,\sum_{b=1}^n q^b p_b \,  \xi \right] - \left[Q_a,\sum_{b=1}^n  p_b P^b \right] + \left[Q_a, \sum_{b=1}^n q^b Q_b \right]\nonumber\\
%						& = - p_a \xi - \sum_{b=1}^n q^b p_b \left[Q_a,\xi \right]  - \sum_{b=1}^n p_b \left[Q_a,P^b \right] + Q_a + \sum_{b=1}^n q^b \left[Q_a,Q_b \right] \nonumber\\
%						& = -p_a \xi + p_a \xi + Q_a \nonumber\\
%	\pounds_{X_{h_S}} Q_a	& = Q_a
%	\end{align}
\beq
\pounds_{X_{h_S}} Q_a	 = Q_a
\eeq
and
	\beq
	\pounds_{X_{h_S}} P^a = - P^a,
	\eeq
while the action on the dual 1-forms is given by
	\begin{align}
	\pounds_{X_{h_S}} \d q^a	& = \iiota_{X_{h_S}} \d \left[\d q^a \right]  + \d \left[\iiota_{X_{h_L}} \d q^a \right]\nonumber\\
						& = \d \left[\d q^a \left(\sum_{b=1}^n q^b p_b \,  \xi +\sum_{b=1}^n \left[ p_b P^b - q^b Q_b \right] \right) \right]\nonumber\\
						%& = \d \left[- q^a \right]\nonumber\\
			\pounds_{X_{h_S}} \d q^a & = -\d q^a
	\end{align}
and
	\beq
	\pounds_{X_{h_S}} \d p_a = \d p_a.
	\eeq
	
Clearly, the generators of Legendre symmetries and polarization scalings do not commute. Indeed,
	\begin{align}
	\left[X_{h_S},X_{h_L} \right] &= \left[ \sum_{a = 1}^n \left\{ p_a q^a \xi - q^a Q_a + p_a P^a \right\}, \sum_{i=1}^m \left\{ \frac{1}{2} (p_i{}^2 + q^i{}^2 )\xi - p_i Q_i + q^i P^i \right\} \right] \nonumber \\
	&= \sum_{i=1}^m \left\{ \left({p_i}^2 - {q^i}^2  \right) \xi - 2 \left(p_i Q_i + q^i P^i \right) \right\}.
	\end{align}

\section{Almost Contact and Almost para-contact Structures}\label{sec:ACS-APCS}

An almost contact structure is a triplet $(\eta,\xi, \phi)$ consisting of a contact 1-form $\eta$, its corresponding Reeb vector field $\xi$ and an automorphism $\phi:T\T \longrightarrow T\T$ such that \cite{blair2010}
	\beq
	\label{eq.ACS}
	\phi^2 = \phi \circ \phi = - \mathbbm{1} + \eta \otimes \xi \quad \text{with} \quad \phi(\xi)=0 \quad \text{and} \quad \eta \circ \phi = 0.
	\eeq
Here, $\mathbbm{1}$ represents the identity map on $T\T$. In this sense, the map $\phi$ corresponds to the extension of an almost complex structure on a symplectic manifold to the contact case. 

Similarly, an almost para-contact structure is a map $\varphi:T\T \longrightarrow T\T$ such that \cite{Bravetti2015} 
	\beq
	\label{eq.APS}
	\varphi^2 = \varphi \circ \varphi = \mathbbm{1} - \eta \otimes \xi \quad \text{with} \quad \varphi(\xi) = 0 \quad \text{and} \quad \eta \circ \varphi = 0.
	\eeq
Notice that this definition is less restrictive that the one in \cite{Zamkovoy2007} in the sense that the latter takes on the eigenvalues of the almost complex structure to classify the sub-bundles $T\mathcal{E}$, while we do not consider such a classification. In the canonical basis it is expressed as
	\beq
	\mathbbm{1} = \eta \otimes \xi + \sum_{i=1}^n \left[\d q^i \otimes Q_i + \d p_i \otimes P^i\right]
	\eeq
from which the basis expressions for $\phi^2$ and $\varphi^2$ follow
	\beq
	\label{eq.phisqr}
	\phi^2 = - \sum_{a=1}^n \left[\d q^a \otimes Q_a + \d p_a \otimes P^a\right] \quad \text{while} \quad \varphi^2 = \sum_{a=1}^n \left[\d q^a \otimes Q_a + \d p_a \otimes P^a\right].
	\eeq
As both transformations act non-trivially only on $\mathcal{D}$, we can express $\phi$ and $\varphi$ as their corresponding actions on the generators of $\mathcal{D}$. 

\subsection{Almost Contact Structure}
\label{sec.ACS}

Let us  begin by examining the almost contact case.  Given a chosen polarization and the canonical almost complex structure $\mathcal{J}:\mathcal{D} \to \mathcal{D}$, $\mathcal{J}^2 = -\mathbbm{1}$ such that $\mathcal{J}(Q_a) = - P^a$ and $\mathcal{J}(P^a) = Q_a$,  a possible form for $\phi$  is  as a map exchanging the polarization for $\mathcal{D}$, that is, 
	\beq\label{eq.contactpol}
		\phi (\xi) 	 = 0, \quad \phi (Q_a)	 = -P^a \quad \text{and} \quad%
		\phi(P^a)	 = Q_a.	
	\eeq
Note that this is merely a $\pi/2$-rotation acting on the generators of the contact distribution and, indeed, successive applications of the transformation yield a rotation by $\pi$, namely
	\begin{align}
	\label{eq.actphi1}
	\phi \left[\phi(\xi) \right]	& = \phi\left[0 \right] = 0\\
	\phi \left[\phi(Q_a) \right]	& = \phi\left[-P^a \right] = -Q_a\\
	\label{eq.actphi2}
	\phi \left[\phi(P^a) \right]	& = \phi\left[Q_a \right] = -P^a.
	\end{align}
Thus, we can express $\phi$ in terms of the basis \eqref{eq.HCM}
	\beq
	\phi = \sum_{a=1}^n \left[\d p_a \otimes Q_a - \d q^a \otimes P^a \right], \label{ACS:structure}
	\eeq
which clearly satisfies \eqref{eq.ACS}, [cf. equations \eqref{eq.actphi1} - \eqref{eq.actphi2} with \eqref{eq.phisqr}].

Such form  suggests that $\phi$  has rotational (Legendre) symmetry. Indeed, using \eqref{eq.lielq} - \eqref{eq.lieldp} one directly obtains
\beq
{\rm\pounds}_{X_{h_L}} \phi 	 =0.
\eeq
 
%	\begin{align}
%	{\rm\pounds}_{X_{h_L}} \phi 	& = \pounds_{X_{h_L}} \sum_{a=1}^n \left[\d p_a \otimes Q^a - \d q^a \otimes P_a \right]\nonumber\\
%							& = \sum_{a=1}^n \left[\pounds_{X_{h_L}} \left(\d p_a \otimes Q^a \right) - \pounds_{X_{h_L}} \left(\d q^a \otimes Q_a\right) \right]\nonumber\\
%							& = \sum_{a=1}^n \left[\pounds_{X_{h_L}} \d p_a \otimes Q_a + \d p_a \otimes \pounds_{X_{h_L}} Q_a 
%							- \pounds_{X_{h_L}} \d q^a \otimes P^a - \d q^a \otimes \pounds_{X_{h_L}} P^a \right]\nonumber\\
	%						& = \sum_{a=1}^n \left[\d q^a \otimes Q_a - \d p_a \otimes P^a + \d p_a \otimes P^a -\d q^a \otimes Q_a  \right]\nonumber\\\label{eq.philegsym}
%							& = 0.
%	\end{align}
It is also easy to see that such a transformation does not exhibit scaling symmetry, that is
	\beq
	\pounds_{X_{h_S}} \phi = 2 \sum_{a=1}^n \left[ dp_a \otimes Q_a + \d q^a \otimes P^a \right].
	\eeq

\subsection{Almost para-contact Structure}\label{sec:APCS}

We here consider three immediately obvious ways of satisfying \eqref{eq.APS}, namely, $\pi$-rotations
	\beq
	\label{eq.acspi}
		\varphi_\pi (\xi) 	 = 0, \quad \varphi_\pi (Q_a)	 = -Q_a \quad \text{and} \quad%
		\varphi_\pi(P^a)	 = -P^a,	
	\eeq
partial polarization reflections
	\beq
	\label{eq.prs}
		\varphi_r (\xi) 	 = 0, \quad \varphi_r (Q_a)	 = Q_a \quad \text{and} \quad%
		\varphi_r(P^a)	 = -P^a,	
	\eeq
and the composition of clockwise $\pi/2$-rotations and partial reflections, $\varphi_s = \varphi_r \circ \varphi_{\pi/2}$ \footnote{The composition in the reverse order is  $\varphi_{\tilde{s}} = \varphi_{\pi/2} \circ \varphi_r  = -\varphi_s$, therefore we will only analyze the automorphism $\varphi_s$.}
	\beq \label{eq.pi2r}
	\varphi_s (\xi) = 0, \quad \varphi_s (Q_a) = P^a \quad \text{and} \quad \varphi_s(P^a) = Q_a.
	\eeq

These three automorphisms satisfy \eqref{eq.APS}, therefore each of them is an almost para-contact structure
	\beq
	\varphi_{\pi}{}^2 = \varphi_r{}^2 = \varphi_s{}^2 = \mathbbm{1} - \eta \otimes \xi.
	\eeq

Their corresponding local expressions are 
	\beq
	\varphi_\pi = - \sum_{a=1}^n \left[\d q^a \otimes Q_a + \d p_a \otimes P^a \right] \label{APCS:StructureRotations}
	\eeq
and
	\beq
	\varphi_r = \sum_{i=a}^n \left[\d q^a \otimes Q_a - \d p_a \otimes P^a \right], \label{APCS:StructureReflections}
	\eeq 
and
	\beq \label{eq.acspi2}
	\varphi_s = \sum_{a=1}^n \left[ \d q^a \otimes P^a + \d p_a \otimes Q_a \right].
	\eeq
It is also straightforward to show that
	\beq
	\label{eq.apcbothsym}
	\pounds_{X_{h_L}}\varphi_\pi = 0, \quad \text{and} \quad \pounds_{X_{h_S}} \varphi_\pi = 0,  
	\eeq
and
	\beq \label{eq.apc.r.sym}
	\pounds_{X_{h_L}}\varphi_r = - 2 \sum_{i=1}^n \left[\d p_i \otimes Q_i + \d q^i \otimes P^i \right] \quad \text{and} \quad \pounds_{X_{h_S}} \varphi_r = 0, 
	\eeq
whilst 
	\beq \label{eq.apc.s.sym1}
	\pounds_{X_{h_L}} \varphi_s = 2 \sum_{i=1}^m \left[ \d q^i \otimes Q_i - \d p_i \otimes P^i \right] \quad \text{and} \quad \pounds_{X_{h_s}} \varphi_s = -2 \sum_{a=1}^n \left[ \d q^a \otimes P^a - \d p_a \otimes Q_a \right].
	\eeq
Nevertheless, it can be shown that
	\beq \label{eq.apc.s.sym2}
	\pounds_{X_{h_L}} \circ \pounds_{X_{h_s}} \varphi_s = \pounds_{X_{h_s}} \circ \pounds_{X_{h_L}} \varphi_s = 0,
	\eeq
therefore, it is trivially satisfied that
	\beq  \label{eq.apc.s.sym3}
	\pounds_{[X_{h_L}, X_{h_s}]} \varphi_s = 0.
	\eeq

As one might have expected, $\varphi_\pi$ has rotation symmetry, and thus it is polarization (Legendre) invariant. Additionally, it also has polarization scaling invariance. The structure $\varphi_s$ is not polarization nor scaling invariant, however it is symmetric under the composition in any order of these two transformations. On the other hand, $\varphi_r$ is not propagated along the flow of the infinitesimal generator of the Legendre transformations. This is not surprising since $\varphi_r$ acts on $\mathcal{D}$ as a change in orientation, which is not obtained by any sequence of rotations. However, it does present polarization scaling symmetry. This latter structure plays a central role in the construction of metric tensors for the various proposals in geometric thermodynamics, as we will now show.

\section{Associated metrics and geometric thermodynamics}\label{sec:Metrics-ACS-APCS}

Most studies in contact geometry build upon their corresponding counterpart in the symplectic case. Thus, it can be seen that almost contact structures (resp. almost para-contact structures) are extensions of  almost complex structures. Thus, an Hermitian metric on an almost complex manifold satisfies
	\beq
	g(\mathcal{J}X,\mathcal{J}Y) = g(X,Y),
	\eeq 
where $X$ and $Y$ are vector fields on a symplectic manifold and $\mathcal{J}$ is an almost complex structure.

A similar structure arises in almost contact manifolds \cite{blair2010} (resp. almost para-contact manifolds \cite{Zamkovoy2007}). We say that a metric tensor is a  \emph{compatible metric} if it satisfies \cite{blair2010,Zamkovoy2007}
	\beq
	\label{eq.asociate}
	g\left(\phi X, \phi Y \right) = g(X,Y) -\eta(X) \eta(Y) \quad \left(\ \text{resp.} \quad g\left(\varphi X, \varphi Y \right) =  - \left[g(X,Y) - \eta(X) \eta(Y)\right]\ \right),
	\eeq
and $(\phi,\xi,\eta, g)$ is called an almost contact metric structure (resp. $(\varphi, \xi, \eta, g)$ is called an almost para-contact metric structure). 	

If the metric for the almost contact (resp. para-contact) metric structure satisfies
	\beq \label{eq.associate}
	g(X,\phi Y) = \d \eta (X,Y) \quad ( \text{resp.} \quad g(X, \varphi Y) = \d \eta (X,Y)\, ) ,
	\eeq
it is said that $g$ is an \emph{associated metric} and $(\phi, \xi, \eta, g)$ a contact metric structure (resp. $(\varphi, \xi, \eta, g)$ a para-contact metric structure).

Given an almost contact structure $(\eta,\xi,\phi)$ (resp. almost para-contact structure $(\eta,\xi,\varphi)$), a metric tensor can be constructed solely from these ingredients. Here, we consider metric tensors in a broader sense, that is,  $(0,2)$ non-degenerate and symmetric tensor fields,  while relaxing the condition of begin positive definite. These metrics so constructed are, for the almost contact structure
	\beq \label{eq.contactmetric}
	g = \eta \otimes \eta + \d \eta \circ \left(\phi\, \otimes\, \mathbbm{1} \right),% \quad \left[\text{resp.}\quad  g = \eta \otimes \eta + \d \eta \circ \left(\varphi \otimes \mathbbm{1} \right)\right].
	\eeq
and for the almost para-contact structure \footnote{The metric with a plus sign, $g = \eta \otimes \eta + \d \eta \circ \left(\varphi\, \otimes\, \mathbbm{1} \right)$, is compatible if $\d \eta \circ \left(\varphi\, \otimes\, \mathbbm{1} \right)$ is symmetric, but it is not an associated metric.}
	\beq \label{eq.paracontactmetric}
	g = \eta \otimes \eta - \d \eta \circ \left(\varphi\, \otimes\, \mathbbm{1} \right),
	\eeq
where it must be understood that
	\beq 
	(\phi \, \otimes \, \mathbbm{1})(X,Y) = (\phi X, Y) \quad (\text{resp.} \quad (\varphi \, \otimes \, \mathbbm{1})(X,Y) = (\varphi X, Y)\,),
	\eeq
for $X,\, Y \in T\mathcal{T}$, i.e. here $\otimes$ acts as a separator for $\phi$ (resp. $\varphi$) and $\mathbbm{1}$ acting on $T_p\mathcal{T} \times T_p\mathcal{T}$ for $p \in \mathcal{T}$.

It must be noticed that $\d \eta \circ (\phi \, \otimes \, \mathbbm{1})$ (resp. $\d \eta \circ (\varphi \, \otimes \, \mathbbm{1})$) does not necessarily yields a metric for any almost contact (para-contact) structure, it must be a symmetric non-degenerate tensor on its restriction to $\mathcal{D}$ to do so.

Then, it is sufficient for the tensor $\d \eta \circ (\phi \, \otimes \, \mathbbm{1})$ (resp. $\d \eta \circ (\varphi \, \otimes \, \mathbbm{1})$) to be symmetric and non-degenerate on $\mathcal{D}$ to render the metric \eqref{eq.contactmetric} (resp. \eqref{eq.paracontactmetric}) a compatible metric with the almost contact (para-contact) structure, i.e., condition \eqref{eq.asociate} is satisfied for $g$. 

Likewise, it can also be shown that if metrics constructed as in \eqref{eq.contactmetric} and \eqref{eq.paracontactmetric} are compatible metrics, then they also are associated metrics. 

\subsection{Associated metric to an Almost Contact Structure}

The differential of the contact 1-form is a $(0,2)$ antisymmetric tensor field, then in \eqref{eq.contactmetric} the use of a contact almost structure can be used to render this tensor symmetric. Let us consider the canonical almost contact structure presented in subsection \ref{sec.ACS}. Its action on a vector field $X\in T\T$ is given by
	\begin{align}
	\phi X 	& = \sum_{a=1}^n \left[\d p_a \otimes Q_a - \d q^a \otimes P^a \right] \left(X^w \xi +\sum_{b=1}^n\left[ X^b_q Q_b + X^b_P P^b\right]\right) \nonumber\\
			& = \sum_{a=1}^n \left[X_P{}_a Q_a - X^Q{}^a P^a \right], 
	\end{align}
and the combination
	\begin{align}
	\d \eta \left(\phi X, Y\right)	& = -\frac{1}{2}\sum_{a=1}^n \left[ \d p_a \otimes \d q^a  - \d q^a \otimes \d p_a \right]\left(\phi X,Y\right) \nonumber\\
						& = -\frac{1}{2}\sum_{a=1}^n \left[\d p_a \left(-X^Q{}^a P_a\right) \d q^a \left(Y^Q{}^a Q_a\right) - \d q^a \left(X_P{}_a Q_a \right)\d p_a \left(Y_P{}_a P^a\right)  \right]\nonumber\\
						& = \frac{1}{2} \sum_{a=1}^n\left[ X^Q{}^a Y^Q{}^a + X_P{}_a Y_P{}_a \right]\nonumber\\
						& = \frac{1}{2}\sum_{a=1}^n \left[\d q^a \otimes \d q^a + \d p_a \otimes \d p_a\right] (X, Y)
	\end{align}
is a symmetric and non-degenerate 2-rank tensor field restricted to $\mathcal{D}$. Therefore, the metric \eqref{eq.contactmetric} with $\phi$ given by \eqref{ACS:structure} is compatible and associated. 

In the local basis \eqref{eq.contactmetric} with \eqref{ACS:structure} is written as
	\beq \label{metric.acs}
	g = \eta \otimes \eta + \frac{1}{2}\sum_{a=1}^n \left[\d q^a \otimes \d q^a + \d p_a \otimes \d p_a\right].
	\eeq
Thus we see that the metric obtained from the canonical almost contact structure renders the frame defined from the Heisenberg group commutation relations \eqref{eq.HCM}, orthogonal. That is
	\beq
	g(\xi,Q_a) = g(\xi,P_a) =g(Q_a,P^b) = 0, \quad g(\xi,\xi) = 1
	\eeq
and
	\beq
	g(Q_a,Q_b) = g(P^a, P^b) =\left\{\begin{array}{ll}
						\frac{1}{2} & a=b\\
						0 & a\neq b
						\end{array} \right. 
	\eeq
	
Albeit the  structure of the metric resembles that of the Euclidean space, it is not the case. The metric is expressed in terms of the dual frame. In particular, a direct calculation of the Ricci curvature tensor yields
	\begin{align}
	\label{eq.etaeinstein}
	{\rm Ric} 	& = 2 n\left(\eta \otimes \eta\right)  - \sum_{a=1}^n \left[\d q^a \otimes \d q^a + \d p_a \otimes \d p_a \right]\nonumber\\
		& = \lambda \eta \otimes \eta + \nu g,  
	\end{align}
where $\lambda = 2n + 2$ and $\nu = -2$ (For a detailed derivation see section 4 equation (92) and (93) of \cite{Bravetti2015}). When condition \eqref{eq.etaeinstein} is satisfied, it is said that $(\eta,\xi,\phi,g)$ is an $\eta$-Einstein manifold.
%Indeed, such metric satisfies \eqref{eq.asociate}

It can be shown that the Lie derivative with respect to a vector field $X \in T\mathcal{T}$ of \eqref{eq.contactmetric} (resp. \eqref{eq.paracontactmetric}) is
	\beq \label{lieder.g}
	\pounds_{X} g = \pounds_X (\eta \otimes \eta) + \left( \pounds_X \d \eta \right) \circ (\phi \, \otimes \, \mathbbm{1}) + \d \eta \circ (\pounds_X \phi \, \otimes \, \mathbbm{1}).
	\eeq
Therefore, the Lie derivative of $g$ with respect to a contact hamiltonian vector field $X_h$ of an horizontal hamiltonian $\xi(h) = 0$, reduces to 
	\beq \label{liederhor.g}
	\pounds_{X_h} g = \d \eta \circ (\pounds_{X_h} \phi \, \otimes \, \mathbbm{1}) \quad (\text{resp.} \quad \pounds_{X_h} g = \d \eta \circ (\pounds_{X_h} \varphi \, \otimes \, \mathbbm{1}) \ ).
	\eeq
Thus, for the transformations generated by $X_h$ to be an isometry of the metrics \eqref{eq.contactmetric} and \eqref{eq.paracontactmetric} it is sufficient to have $\pounds_{X_h} \phi = 0$ and $\pounds_{X_h} \varphi = 0$, respectively. That is, the symmetry property of the almost contact (para-contact) structure is inherited to the metrics here considered. 

Therefore, we have that \eqref{metric.acs} has Legendre transformations as symmetries, i.e.
	\beq
	\pounds_{X_{h_L}} g = 0,
	\eeq
while it is clear that it does not have polarization scaling transformations as an isometry
	\beq
	\pounds_{X_{h_S}} g=  \sum_{a=1}^n  \left[ \d p_a \otimes \d p_a - \d q^a \otimes \d q^a\right] .
	\eeq
%Since the metric tensor \eqref{metric.acs} is constructed from the almost contact structure \eqref{eq.contactpol} satisfying the Legendre symmetry condition \eqref{eq.philegsym}, it follows that
%	\begin{align}
%	{\rm\pounds}_{X_{h_L}} g 	& =\sum_{i=1}^n \pounds_{X_{h_L}}  \left[   \eta \otimes \eta + \frac{1}{2} \left(\d q^i \otimes \d q^i + \d p_i \otimes \d p_i \right)  \right]\nonumber\\
%							& = \frac{1}{2} \sum_{i=1}^n \left[\pounds_{X_{h_L}} \left(\d q^i \otimes \d q^i \right) + \pounds_{X_{h_L}} \left(   \d p_i \otimes \d p_i  \right) \right]\nonumber\\
%							& = \sum_{i=1}^n \left[\pounds_{X_{h_L}} \d q^i \otimes \d q^i + \d q^i \otimes \pounds_{X_{h_L}} \d q^i
%							+ \pounds_{X_{h_L}} \d p_i \otimes \d p_i +  \d p_i \otimes \pounds_{X_{h_L}} \d p_i \right]\nonumber\\
%							& = 0.
%	\end{align}
%Thus, this metric shares the Legendre symmetry of the contact 1-form. 
%However, it is also clear that it does not share its polarization scaling symmetry, 
%	\beq
%	\pounds_{X_{h_S}} g=  \sum_{i=1}^n  \left[ \d p_i \otimes \d p_i - \d q^i \otimes \d q^i\right] .
%	\eeq

\subsection{Associated metric to an almost para-contact Structure}

In this subsection we consider the almost-para-contact structures explored in subsection \ref{sec:APCS}, namely those related to $\pi$-rotations, to partial polarization reflections and to the composition of $\pi/2$-rotations with the partial polarization reflections.

 The action of $\varphi_{\pi}$ on a vector field $X\in T\T$ is given by
	\begin{align}
	\varphi_{\pi} X 	& = - \sum_{a=1}^n \left[\d q^a \otimes Q_a + \d p_a \otimes P^a \right] \left(X^w \xi +\sum_{b=1}^n\left[ X^Q{}^b Q_b + X_P{}_b P^b\right]\right) \nonumber\\
			& = -\sum_{a=1}^n \left[X_P{}_a P^a + X^Q{}^a Q_a \right],
	\end{align}
and the combination
\beq
\d \eta \left(\varphi_{\pi} X, Y\right)	 =\frac{1}{2}\sum_{a=1}^n \left[\d p_a \otimes \d q^a - \d q^a \otimes \d p_a\right] (X, Y)
\eeq

	\begin{align}
	\d \eta \left(\varphi_{\pi} X, Y\right)	& = -\frac{1}{2}\sum_{a=1}^n \left[ \d p_a \otimes \d q^a  - \d q^a \otimes \d p_a \right]\left(\varphi_{\pi} X,Y\right) \nonumber\\
						& = -\frac{1}{2}\sum_{a=1}^n \left[\d p_a\left(-X_P{}_a P^a\right) \d q^a \left(Y^Q{}^a Q_a\right) - \d q^a \left(-X^Q{}^a Q_a \right)\d p_a \left(Y_P{}_a P^a\right)  \right]\nonumber\\
						& = \frac{1}{2} \sum_{a=1}^n\left[ X_P{}_a Y^Q{}^a - X^Q{}^a Y_P{}_a\right]\nonumber\\
						& = \frac{1}{2}\sum_{a=1}^n \left[\d p_a \otimes \d q^a - \d q^a \otimes \d p_a\right] (X, Y)
	\end{align}
is antisymmetric. Then, it is not possible to construct a metric as in the almost contact case, instead this almost para-contact structure can be used, in analogy, to define a 2-rank covariant tensor field given by
	\beq \label{eq.paracotactmetric1}
	\alpha_{\pi} = \eta \otimes \eta + \d \eta \circ \left(\varphi_{\pi} \otimes \mathbbm{1} \right)  = \eta \otimes \eta - \d \eta,% \quad \left[\text{resp.}\quad  g = \eta \otimes \eta + \d \eta \circ \left(\varphi \otimes \mathbbm{1} \right)\right].
	\eeq
which in the local basis is written as
	\beq
	\alpha_{\pi} = \eta \otimes \eta + \frac{1}{2}\sum_{a=1}^n \left[\d p_a \otimes \d q^a - \d q^a \otimes \d p_a\right].
	\eeq
	
	Since the  tensor \eqref{eq.paracotactmetric1} is constructed from the almost-para-contact structure \eqref{eq.acspi} with rotational and polarization scaling symmetries [cf. equation \eqref{eq.apcbothsym}], $\alpha_{\pi}$ satisfies
	\beq
	\pounds_{X_{h_L}} \alpha_{\pi}=0 \quad \text{and} \quad \pounds_{X_{h_S}} \alpha_{\pi}=0.
	\eeq	

Similarly, the metric constructed from the almost para-contact structure with polarization reflection symmetry [cf. equation \eqref{eq.prs}] can be constructed  by considering the action of $\varphi _r$ on a vector field $X\in T\mathcal{T}$, 
	\begin{align}
	\varphi_{r} X 	& = \sum_{a=1}^n \left[\d q^a \otimes Q_a - \d p_a \otimes P^a \right] \left(X^w \xi +\sum_{b=1}^n\left[ X^Q{}^b Q_b + X_P{}_b P^b\right]\right) \nonumber\\
			& = \sum_{a=1}^n \left[X^Q{}^a Q_a - X_P{}_a P^a  \right]
	\end{align}
and, again, noting that
\beq
\d \eta \left(\varphi_{r} X, Y\right)=\frac{1}{2}\sum_{a= 1}^n \left[\d p_a \otimes \d q^a + \d q^a \otimes \d p_a\right] (X, Y)
\eeq
is symmetric and non-degenerate on $\mathcal{D}$. Therefore, the metric on $\mathcal{T}$ constructed as \eqref{eq.paracontactmetric} with $\varphi = \varphi_r$
	\beq  \label{eq.metric.r}
	g_r = \eta \otimes \eta - \d \eta \circ (\varphi_r \, \otimes \, \mathbbm{1})
	\eeq
is compatible and consequently associated too. Expressed in terms of the local basis takes the following form
	\beq  \label{eq.metric.r.basis}
	g_{r} = \eta \otimes \eta - \frac{1}{2}\sum_{a=1}^n \left[\d p_a \otimes \d q^a + \d q^a \otimes \d p_a\right].
	\eeq
Thus, the metric $g_{r}$ does not render the frame \eqref{eq.HCM} orthonormal, instead,
	\beq
	g_{r}(\xi,Q_a) = g_r(\xi,P_a) = 0, \quad g_{r}(\xi,\xi) = 1
	\eeq
and
	\beq
	g_{r}(Q_a,Q_b) = g_{r}(P^a, P^b) =0, \quad g_{r}(P^a,Q_b)=g_{r}(Q_b,P^a) \, =- \frac{1}{2} \delta^a_b .
	\eeq

From \eqref{eq.apc.r.sym} we deduce that the metric $g_r$ has scaling symmetry 	
	\beq
	\pounds_{X_{h_S}} g_{r} = 0,
	\eeq	
but they fail to be propagated along the Legendre symmetry generator, ie.
	\beq \label{nonlegendresymmetry}
	\pounds_{X_{h_L}} g_{r}= - \sum_{i=1}^m \left[\d q^i \otimes \d q^i - \d p_i \otimes \d p_i \right] .
	\eeq

For the metric constructed from $\varphi_s$ we follow the same steps as before. The action of $\varphi_s$ on $X \in T\mathcal{T}$ is
	\begin{align}
	\varphi_s X &= \sum_{a=1}^n \left[\d q^a \otimes P^a + \d p_a \otimes Q_a \right] \left(X^w \xi + \sum_{b=1}^n \left[ X^Q{}^b Q_b + X_P{}_b P^b \right] \right), \nonumber \\
	&= \sum_{a=1}^n \left[ X^Q{}^a P^a + X_P{}_a Q_a \right],
	\end{align}
and 
\beq
\d \eta (\varphi_s X, Y) =-\frac{1}{2} \sum_{a=1}^n \left[ \d q^a \otimes \d q^a - \d p_a \otimes \d p_a \right](X,Y)
\eeq
is a symmetric non-degenerate tensor on $\mathcal{D}$. Hence the metric constructed as \eqref{eq.paracontactmetric} from $\varphi_s$ is a compatible and associated metric, which in terms of the local basis takes the following form
	\beq \label{eq.met.s.basis}
	g_s = \eta \otimes \eta + \frac{1}{2} \sum_{a=1}^n \left[ \d q^a \otimes \d q^a - \d p_a \otimes \d p_a \right].
	\eeq
The frame \eqref{eq.HCM} is pseudo-orthogonal with respect to $g_s$
	\beq
	g_s(\xi,Q_a) = g_s(\xi,P_a) =g_s(Q_a,P^b) = 0, \quad g_s(\xi,\xi) = 1
	\eeq
and
	\beq
	g_s(Q_a,Q_b) = \begin{cases}
					\frac{1}{2} & a = b \\
					0 & a\neq  b
					\end{cases} \quad \text{and} \quad  g_s(P^a, P^b) = \begin{cases} -\frac{1}{2} & a = b \\
					0 & a\neq b .
					\end{cases}
	\eeq
In what refers to the Legendre and scaling transformations, given the properties of $\varphi_s$, these are not isometries for this metric, instead
	\beq
	\pounds_{X_{h_L}} g_s = - \sum_{i=1}^m \left( \d q^i \otimes \d p_i + \d p_i \otimes \d q^i \right) \quad \text{and} \quad \pounds_{X_{h_s}} g_s = - \sum_{a=1}^n \left( \d p_a \otimes \d p_a + \d q^a \otimes \d q^a \right),
	\eeq
while it is symmetric for the composition of the transformations in any order
	\beq
	\pounds_{X_{h_L}} \circ \pounds_{X_{h_s}} g_s = \pounds_{X_{h_s}} \circ \pounds_{X_{h_L}} g_s = 0 \quad \text{then} \quad \pounds_{[X_{h_L}, X_{h_s}]} g_s = 0,
	\eeq
just as $\varphi_s$ is. 

Equation \eqref{nonlegendresymmetry} shows us that \eqref{eq.metric.r} is not Legendre invariant. The pullback of this metric onto its corresponding Legrendre submanifold takes the form of a Hessian metric, i.e. the components of the metric form a Hessian matrix, and plays a preponderant part in the geometric description of thermodynamics \cite{Weinhold1975, 1979Ruppeiner}. It has been shown that such a metric carries information about the fluctuations around equilibrium \cite{1995Rupp}. This metric does not preserve, in general, its Hessian form under a change of thermodynamic potential \cite{1984Salamon} making the geometric description of fluctuations ensemble dependent \cite{termometrica}. In \cite{Salamon83} it was investigated a group of transformations that leave Weinhold's metric invariant, finding that a total Legendre transformation preserves the form of the metric while all the partial Legendre transformation do not belong to this group. Other attempts to find metrics for the space of thermodynamic equilibrium states which are invarianat under Legendre transformations were conducted in a strongly dependent thermodynamic coordinate framework \cite{2007Quev} yielding  a set of metrics whose components are proportional to the components of the Hessian metrics. Recently, it was shown in \cite{reparametrizations, Pineda2019} that Legendre invariant metrics can be related physically to reparametrizations of the thermodynamic state variables. The set of metrics found in \cite{Pineda2019} have as a particular case those of \cite{2007Quev}. Hence, it is physically relevant and mathematically consistent  to generate Legendre invariant metrics from purely geometric structures. 

In the next section we construct a family of such metrics from an automorphism which is a modification to the almost para-contact structure related to partial polarization reflections $\varphi_r$. This automorphism is not an almost para-contact structure but it will be shown that it satisfy some generalized version of \eqref{eq.APS}.

\section{Polarization independent metrics}\label{sec:CRA}

Let $\psi: \mathcal{E} \to \mathcal{T}$ be an embedding of  $\mathcal{E}$ into the contact manifold $\mathcal{T}$ defined by the condition $\psi^* \eta = 0$ and $\mathrm{dim}(\mathcal{E}) = n$. In local coordinates this embedding is defined as 	\beq
	(q^a) \mapsto \left( q^a,w=\bar{w}(q^a), p_a=\frac{\partial}{\partial q^a} \bar{w}\right).
	\eeq
In geometric thermodynamics, the Legendre submanifold $\mathcal{E}$ is referred  as the \emph{space of equilibrium states} since its defining  condition can be  identified with the first law of thermodynamics [cf. equation \eqref{eq.eta1}, above]. In addition, it is straightforward to equip such a sub-manifold with a Riemannian structure by considering the induced metric
	\beq
	\psi^* g_r = -\frac{\partial^2 \bar{w}}{\partial q^a \partial q^b} \d q^a \otimes \d q^b.
	\eeq

This is a widely used metric in the geometric description of thermodynamics, where the function $\bar{w}(q^a)$ is usually identified with the entropy or the internal energy of the thermodynamic system \cite{1979Ruppeiner, Weinhold1975}.

On the contact manifold, different polarizations  correspond to distinct choices of thermodynamic potentials (cf. section \ref{sec:Heiseberg}, above). In this section we aim to construct polarization independent metrics on $\mathcal{T}$, i.e. metrics which have Legendre transformations as isometries, subject to the condition that the components of its pullback onto its corresponding Legendre submanifold are proportional to the components of the Hessian of a thermodynamic potential. We have seen that the contact form is polarization independent, that translates in the thermodynamic language into the Legendre invariance of the description of equilibrium thermodynamics, namely, the Legendre invariance of the first law. It has been argued that the metric on the space of equilibrium states with components given by the negative  Hessian of the entropy encloses the information of the second law and its positive definiteness accounts for the local stability conditions of the system \cite{1995Rupp}. It is also well known that such a Hessian metric  does not posses a  symmetry under a general Legendre transformation, we can see precisely that from equation \eqref{nonlegendresymmetry}.  As we will see it will be necessary to abandon  the possibility of having a metric contact manifold. Nevertheless, a generalization of the notion of metric contact manifold can be proposed in terms of a modification of the almost para-contact structure $\varphi_r$ from which a metric can be constructed which allows the possibility of being Legendre invariant. 

Let us consider an automorphism $\varphi_\Lambda$ constructed from the almost para-contact structure \eqref{APCS:StructureRotations} multiplying each component by a function $\Lambda_{(a)} = \Lambda_{(a)} (w, q,p)$, where $q = (q^1, \ldots, q^n)$ and similarly for $p=p(p_1,\ldots,p_n)$ 
\beq \label{conformalstructe}
\varphi_\Lambda = \sum_{a=1}^n \Lambda_{(a)} \left( \d q^a \otimes  Q_a -  \d p_a \otimes  P^a \right).
\eeq

This tensor field is not a almost contact or para-contact structure, but satisfies 
	\beq  \label{not.structure}
	\varphi_\Lambda (\xi) = 0, \quad \varphi_\Lambda(Q_a) = \Lambda_{(a)} Q_a \quad \text{and} \quad \varphi_\Lambda (P^a) = -\Lambda_{(a)} P^a,
	\eeq
and 
	\beq \label{not.structure2}
	\varphi_\Lambda{}^2 (\xi) = 0, \quad \varphi_\Lambda{}^2 (Q_a) = \Lambda_{(a)}{}^2 Q_a \quad \text{and} \quad \varphi_\Lambda{}^2 (P^a) = \Lambda_{(a)}{}^2 P^a,
	\eeq
which can be expressed as
	\beq  \label{not.structure3}
	\varphi_\Lambda{}^2 = \mathbbm{1}_{\Lambda} - \eta \otimes \xi,
	\eeq
where 
	\beq \label{1omega}
	\mathbbm{1}_{\Lambda} = \eta \otimes \xi + \sum_{a=1}^n \Lambda_{(a)} \left(\d q^a \otimes Q_a + \d p_a \otimes P^a \right).
	\eeq

As we have seen $\varphi_r$ is invariant under polarization scaling but fails to be symmetric under Legendre transformations. Let us analyze the symmetry properties of \eqref{conformalstructe} under these transformations. Let us consider the infinitesimal action of $X_{h_s}$ on $\varphi_\Lambda$ first
	\begin{align}
	\pounds_{X_{h_s}} \varphi_{\Lambda} &= \sum_{a=1}^n \left( \pounds_{X_{h_s}} \Lambda_{(a)} \right) \left( \d q^a \otimes Q_a - \d p_a \otimes P^a \right) + \Lambda_{(a)} \pounds_{X_{h_s}}\left( \d q^a \otimes Q_a - \d p_a \otimes P^a \right)  \nonumber \\
	&= \sum_{a=1}^n \left( \pounds_{X_{h_s}} \Lambda_{(a)} \right) \left( \d q^a \otimes Q_a - \d p_a \otimes P^a \right) .
	\end{align}
Then, for $\varphi_{\Lambda_{(a)}}$ to be polarization scaling invariant it must be satisfied that
	\beq
	\pounds_{X_{h_s}} \Lambda_{(a)} = 0,
	\eeq
and from the expression for the Hamiltonian vector field generating such transformations \eqref{eq.polscaling.X} we obtain the following condition
	\beq \label{Omega.cond}
	\sum_{b=1}^n \left( p_b \frac{\partial \Lambda_{(a)}}{\partial p_b} - q^b \frac{\partial \Lambda_{(a)}}{\partial q^b} \right) = 0 .
	\eeq

The solutions that make $\varphi_\Lambda$ polarization scaling invariant are a set of functions which have the general form
	\beq  \label{Omega.cond1}
	\Lambda_{(a)} = \Lambda_{(a)}\left(w, q^{\tilde{b}}p_{\tilde{b}}, q^{\tilde{b}} p_c, \frac{q^c}{q^{\tilde{b}}}\right),
	\eeq
for all $a = 1, \ldots, n$, where $\tilde{b} \neq c$ is fixed at any value from 1 to $n$ and $c$ takes all the other values. 

The metric constructed from this automorphism \eqref{conformalstructe}
	\beq
	g_{\Lambda} = \eta \otimes \eta - \d \eta \circ (\varphi_{\Lambda} \, \otimes \, \mathbbm{1}),
	\eeq
has the same polarization scaling invariance as $\varphi_\Lambda$ as we have seen before and in the basis \eqref{eq.bracketgenerating} is written as
	\beq  \label{metric.Omega}
	g_{\Lambda} = \eta \otimes \eta - \frac{1}{2} \sum_{a=1}^n \Lambda_{(a)} \left( \d p_a \otimes \d q^a + \d q^a \otimes \d p_a \right).
	\eeq
These metrics are not compatible nor associated because $\varphi_\Lambda$ is not an almost contact or para-contact structure. Let us analyze the behavior of \eqref{metric.Omega} under an infinitesimal Legendre transformation. Its Lie derivative with respect to $X_{h_L}$,
	\beq 
	\pounds_{X_{h_L}} g_\Lambda = - \frac{1}{2} \sum_{i=1}^m \left[ (\pounds_{X_{h_L}} \Lambda_{(i)}) \left( \d p_i \otimes \d q^i + \d q^i \otimes \d p_i \right) + \Lambda_{(i)} \left( \d q^i \otimes \d q^i - \d p_i \otimes \d p_i \right) \right],
	\eeq
shows that the metric is not invariant under an infinitesimal strict contactomorphism as there are not functions $\Lambda_{(i)}$ such that $\pounds_{X_{h_L}} g_\Lambda = 0$.  

Nevertheless, in thermodynamics one is interested in the finite version of the strict contactomorphisms given by \eqref{legendre.transf}, those representing a partial change of the contact polarization in $\mathcal{D}$ as $\pi/2$-rotations of some polarization planes. For $g_\Lambda$ to be Legendre invariant, the functions $\Lambda_{(a)}$ must be such that
	\beq
	\left[{\Phi^m_{\frac{\pi}{2}}}\right]^* g_\Lambda = g_\Lambda.
	\eeq 
Then, we can establish the conditions on the functions $\Lambda_{(a)}$ as follows
	\begin{multline} 
0 =	\left[{\Phi^m_{\frac{\pi}{2}}}\right]^* g_\Lambda - g_\Lambda = \frac{1}{2} \sum_{i=1}^m \left( \left[{\Phi^m_{\frac{\pi}{2}}}\right]^* \Lambda_{(i)} + \Lambda_{(i)} \right) \left( \d q^i \otimes \d p_i + \d p_i \otimes \d q^i \right) \\
	-\frac{1}{2} \sum_{I=m+1}^n \left( \left[{\Phi^m_{\frac{\pi}{2}}}\right]^* \Lambda_{(I)} - \Lambda_{(I)} \right) \left( \d q^I \otimes \d p_I + \d p_I \otimes \d q^I \right).
	\end{multline}

Therefore, for $g_\Lambda$ to be Legendre invariant it must be satisfied that under a Legendre transformation on the $i = 1, \ldots, m$ directions
	\beq
	\left[{\Phi^m_{\frac{\pi}{2}}}\right]^* \Lambda_{(i)} = - \Lambda_{(i)} \quad \text{and} \quad \left[{\Phi^m_{\frac{\pi}{2}}}\right]^* \Lambda_{(I)} = \Lambda_{(I)},
	\eeq
for each $i= 1,\ldots, m$ and $I= m+1,\ldots, n$. 

It is clear that the above conditions impose further restrictions on the functions than those in \eqref{Omega.cond1}. Evidently, we must have $\partial_w \Lambda_{(a)} = 0$. We are interested in having a metric that is invariant under all the Legendre transformations, that is, for any subdivision of the indices $i$ and $I$. Therefore an obvious choice for the functions $\Lambda_{(a)}$ which is a particular case of \eqref{Omega.cond1}, and consequently invariant also under polarization scaling, is 
	\beq
	\Lambda_{(a)} = \Lambda_{(a)}( q^a p_a) 
	\eeq
for $a= 1,\ldots n$. Hence, $\Lambda_{(a)}$ for each $a$ is a function of only the product of the corresponding variables, e.g. $\Lambda_{(1)} = \Lambda_{(1)}(q^1 p_1)$. These functions must also be odd 
	\beq
	\Lambda_{(a)} (- q^a p_a) = - \Lambda_{(a)} (q^a p_a).
	\eeq
It is worth mentioning that there is another simple choice for $\Lambda_{(a)}$ that makes $g_\Lambda$ Legendre invariant, although not invariant under polarization scaling. It was explored in \cite{Pineda2019} from a different approach and considers the functions $\Lambda_{(a)}$ as the product of two functions. That is, $\Lambda_{(a)} = f_{(a)}(q^a) g_{(a)}(p_a)$ where $f$ and $g$ have the same functional form and are odd functions.

In general, given a contact structure and an associated metric, the explicit form of the almost para-contact structure can be obtained by the covariant derivative of the Reeb vector \cite{Zamkovoy2007}, 
\beq \label{paracontstructure1}
\nabla \xi = \varphi -\varphi \kappa\,,
\eeq
 where 
 \beq
 \kappa=\frac{1}{2}\pounds_{\xi}\varphi \,.
 \eeq
The tensor field $\kappa$ vanishes if $\xi$ corresponds to a Killing vector of the associated metric and $\nabla$ is the Levi-Civita connection. Therefore, the almost para-contact structure is defined by
\beq \label{paracontstructure2}
\nabla \xi = \varphi .
\eeq

Since the metric components of \eqref{metric.Omega} does not depend of $\omega$, $\xi$ is indeed a Killing vector of $g_{\Lambda}$. The covariant derivative associated to \eqref{metric.Omega}, of the Reeb vector in the basis $e_{(i)}=\{ \xi , \frac{\partial}{\partial p_a}, \frac{\partial}{\partial q^a} \}$ and dual basis $\theta^{(a)}=\{ \d\omega , \d p_a, \d q^a \}$  is 
\begin{eqnarray}
\nabla \xi &=&\sum_{a,b=0}^{2n}\Gamma^a_{\omega b}\, e_{(a)} \otimes \theta ^{(b)}\nonumber\\
&=& \sum _{a=1}^n \left( \Gamma^{p_a}_{\omega p_a} \frac{\partial}{\partial p_a} \otimes \d p_a + \Gamma^{q^a}_{\omega q^a}\frac{\partial}{\partial q^a} \otimes \d q^a + \Gamma^{\omega}_{\omega q^a}\xi \otimes \d q^a \right)\nonumber \\
&=& - \sum_{a=1}^n \frac{1}{\Lambda_{(a)}} \left[ \d q^a \otimes   Q_{(a)} -  \d p_a \otimes P^a \right] \nonumber\\
&=& - \bar\varphi_\Lambda\,  \label{barconformalstructure}
\end{eqnarray}
where
	\beq \label{phi.barOmega}
	\varphi_{\bar{\Lambda}} = \sum_{a=1}^n \frac{1}{\Lambda_{(a)}} \left[ \d q^a \otimes   Q_{(a)} -  \d p_a \otimes P^a \right].
	\eeq
This confirms that the automorphism \eqref{conformalstructe} is not an almost para-contact structure as we have seen before. Indeed, $-\varphi_{\bar{\Lambda}}$ is significantly different to $\varphi_{\Lambda}$. We can construct a metric from the automorphism given by \eqref{phi.barOmega} as
	\beq\label{barmetric}
	\bar g_\Lambda = \eta \otimes \eta - \d \eta \circ \left(\bar\varphi_{\Lambda} \otimes \mathbbm{1} \right)
	\eeq
and, as we did before, it is now possible to calculate the covariant derivative of the Reeb vector, this time associated to the affine connection of \eqref{barmetric} instead
\begin{eqnarray}
\bar \nabla \xi &=&\sum_{a,b=0}^{2n}\bar \Gamma^a_{\omega b} e_{(a)}\otimes \theta ^{(b)}\nonumber\\
&=& \sum _{a=1}^n \left( \bar \Gamma^{p_a}_{\omega p_a}\frac{\partial}{\partial p_a} \otimes \d p_a + \bar \Gamma^{q^a}_{\omega q^a}\frac{\partial}{\partial q^a} \otimes \d q^a + \bar \Gamma^{\omega}_{\omega q^a}\xi \otimes \d q^a \right)\nonumber \\
&=& \sum_{a=1}^n \Lambda_{(a)} \left[ \d q^a \otimes  Q_{(a)} - \d p_a \otimes P^a \otimes \right]  \nonumber\\
&=& - \varphi_\Lambda. 
\end{eqnarray}

Thus, we observe that the covariant derivative of the Reeb vector with respect to the Levi-Civita connection of the metrics $g_\Lambda$ and $g_{\bar{\Lambda}}$ establish a sort of dual relation between the automorphisms $\varphi_\Lambda$ and $\varphi_{\bar{\Lambda}}$. Hence the introduction of these $\Lambda_{(a)}$ functions multiplying each component of $\varphi_r$ changes the usual properties of the Reeb vector field.

We can conclude that in order to construct a Legendre invariant metric which induces a metric on the Legendre submanifold with components proportional to the Hessian of a function we must give up  the contact metric structure. It is interesting that the modifications proposed allow us to introduce a  further relationship between structures for a metric contact manifold. Indeed,  these \emph{dual} automorphisms satisfy
	\beq \label{paracontactproperty}
	\varphi_\Lambda \circ \varphi_{\bar{\Lambda}} = \varphi_{\bar{\Lambda}} \circ \varphi_\Lambda = \sum_{i=1}^n \left[\d q^i \otimes Q_i + \d p_i \otimes P^i\right] = \mathbbm{1} - \eta \otimes \xi.
	\eeq	
Therefore, these modifications in \eqref{conformalstructe} not only generate families of Legendre invariant metrics, they generate structures with similar, but more general, properties of almost para-contact structures.

Let us close this section by considering the potential significance of this latter class of metrics in the context of contact geometric thermodynamics. Albeit a particular instance of \eqref{metric.Omega} has been used to explore geodesics on a specific Legendre submanifold \cite{2008quevedo},  thermodynamic processes - in general -  lie on $\mathcal{T}$. In this sense  the notion of  thermodynamic processes corresponds to horizontal curves in the thermodynamic phase space (cf. section 5.3 in \cite{HamThermo}), namely, those whose tangent vector at each point is an element of the contact distribution $\mathcal{D}$. However, the length of such curves is yet to be understood. It is expected that such length  should not depend on the contact polarization used along the process. The metric structures \eqref{metric.Omega} and \eqref{barmetric} provide us with  a tool for further exploration in this direction. Moreover, such analysis would allow us to probe geometric thermodynamics beyond the metric Legendre submanifolds of  Ruppeiner \cite{1995Rupp} and Weinhold \cite{Weinhold1975} representations and will be explored in a forthcoming manuscript.

\begin{table}
\begin{center}
\resizebox{1.0\textwidth}{!}{%
\begin{tabular}{| c | c | c | }
\hline
Almost (para) & $\pounds_{X_{h_L}} g$	&	$\pounds_{X_{h_s}}g$	\\ 
contact structure & &  \\ \hline

$\phi$	 &	 0	&	$  \d p_a \otimes \d p_a - \d q^a \otimes \d q^a$ 	 \\

$\varphi_{\pi}$ &  0	&	0 \\

$\varphi_{r}$ &  $ \d q^i \otimes \d q^i - \d p_i \otimes \d p_i $	&	0	\\

$\varphi_s$	&	$-\left( \d q^i \otimes \d p_i + \d p_i \otimes \d q^i \right)$	&	$-\left( \d p_a \otimes \d p_a + \d q^a \otimes \d q^a \right)$	\\

$\varphi_{\Lambda}$	&	 $- \frac{1}{2}  \left[ (\pounds_{X_{h_L}} \Lambda_{(i)}) \left( \d p_i \otimes \d q^i + \d q^i \otimes \d p_i \right)+ \Lambda_{(i)} \left( \d q^i \otimes \d q^i - \d p_i \otimes \d p_i \right) \right]$	&	0	 \\

$\varphi_{\bar{\Lambda}}$	 &	$- \frac{1}{2}  \left[ (\pounds_{X_{h_L}}\frac{1}{ \Lambda_{(i)}}) \left( \d p_i \otimes \d q^i + \d q^i \otimes \d p_i \right)+ \frac{1}{ \Lambda_{(i)}} \left( \d q^i \otimes \d q^i - \d p_i \otimes \d p_i \right) \right]$		&	0	\\

 \hline
\end{tabular}%
}
\label{tab:Table}
\caption{The metric $g$ associated to an almost contact structures $\phi$ is computed from \eqref{eq.contactmetric}. Metrics $g$ associated to almost paracontact structures $\varphi_{r}$ and $\varphi_{s}$ are computed from \eqref{eq.paracontactmetric}. Metrics $g$ of structures $\varphi_{\Lambda}$ and $\varphi_{\bar{\Lambda}}$ are computed from \eqref{eq.paracontactmetric} but are not associated metrics. The $(0,2)$-tensor of the almost para-contact structure $\varphi_{\pi}$ given by \eqref{eq.contactmetric} it is not a metric.} 
\end{center}
\end{table}

\section{Closing remarks}\label{sec:conclusions}

We have analyzed the construction of \emph{associated metrics} using different almost contact and para-contact structures which where defined according to its symmetry properties under change of contact polarizations and polarization scalings. As it is already known, the associated metric constructed from the almost para-contact structure \eqref{APCS:StructureReflections} plays an important role in the geometric description of thermodynamics as it gives a Hessian metric on the Legendre submanifold, which physically represents the space of equilibrium states. We showed that this metric $g_r$ is not invariant under changes of polarization, namely, Legendre transformations, but it has scaling symmetry. This proves that it is not possible to construct an \emph{associated metric} satisfying both properties,  being \emph{Legendre invariant} and  inducing a Hessian metric into  a Legendre submanifold. Although there are  associated metrics to an almost contact structure \eqref{metric.acs} invariant under Legendre transformations, they do not satisfy the second requirement expressed above. We have seen that both -- polarization and scaling invariance -- cannot be simultaneous isometries of an associated metric. 

It was our aim  to construct a metric satisfying a relaxed version of the requirements mentioned above: to be Legendre invariant and to induce a metric on the Legendre manifold whose components are each proportional to the entries of a Hessian matrix. In order to achieve this, we have found that it is necessary to generalize the concept of almost para-contact structures and consequently the notion of a metric contact manifold. This generalization has been established in the last section and it can be understood as an anisotropic scaling of the almost para-contact structure \eqref{APCS:StructureReflections}. The generalized structures that we have found seem to have interesting properties which demand  further exploration.

The Legendre invariant metrics  presented here were constructed with the sole objective of minimally modifying the structures defining a contact metric manifold. Some other works had reached equivalent results using as a motivation some physical criteria \cite{2007Quev, reparametrizations,Pineda2019} and from a different perspective. For instance,  in \cite{Salamon83} it was studied the group of transformations that leave the Hessian metrics invariant. 

We leave for  future work the analysis of the generalized structures found here and their possible consequences beyond the realm of thermodynamics. In particular, let us close this work by noting that a choice of contact polarization for the contact distribution is equivalent to an election of thermodynamic potential. In this sense, a  Legendre transformation is nothing but a mere  change of contact polarization in the context of contact geometry.

\section*{Acknowledgements}

The authors are thankful to JJ for pointing the idea of symplectic polarizations. The authors want to acknowledge the anonymous referees for their insightful suggestions.

\section*{Appendix I. Contact polarization in thermodynamics}

Consider contact manifold $(\mathcal{T},\mathcal{D})$ together with a representative contact 1-form $\eta$, i.e. ${\rm ker}(\eta) = \mathcal{D}$. Let $\mathcal{U}_U,\mathcal{U}_F \subset \mathcal{T}$ be coordinate neighborhoods around a point $q\in\mathcal{T}$ with coordinates $(S,V,U,T,P)$ and $(T,V,F,S,P)$, respectively, such that the contact 1-form is written as
	\beq
	\label{eq.coord1form}
	{\Psi^{-1}}^*(\eta) = \d U - T\d S + P \d V  \quad \text{and} \quad  {\Phi^{-1}}^*(\eta) = d F + S \d T + P \d V,
	\eeq  
where $\Psi:\mathcal{T} \longrightarrow \mathcal{U}_U\subset\mathbb{R}^5 $ and $\Phi:\mathcal{T} \longrightarrow \mathcal{U}_F\subset\mathbb{R}^5 $ are the corresponding coordinate maps. From the definition of a differentiable manifold, the map $\phi^S_{\frac{\pi}{2}}$ is a diffeomorphism between open sets of $\mathbb{R}^5$, inducing the map $\left[\phi^S_{\frac{\pi}{2}} \right]^*$, relating the 1-forms \eqref{eq.coord1form} as
	\beq
	 \left[\phi^S_{\frac{\pi}{2}} \right]^*{\Phi^{-1}}^*(\eta) = {\Psi^{-1}}^*(\eta),
	\eeq
namely
	\beq
	 \left[\phi^S_{\frac{\pi}{2}} \right]^*(d F + S \d T + P \d V) =  \d U - T\d S + P \d V.
	\eeq

Legendre sub-manifolds of $(\mathcal{T},\mathcal{D})$ are integral sub-manifolds  $\varphi:\mathcal{E} \longrightarrow \mathcal{T}$ of maximal dimension defined by the condition $\varphi^*(\eta) = 0$. 

Since the coordinates for each patch $\mathcal{U}_U$ and $\mathcal{U}_F$ are ordered, this yields different embedded sub-manifolds, whose coordinate representation is expressed through the composition
	\beq
	\Psi \circ \ell_U \circ \psi^{-1}: (S,V) \mapsto \left(S, V,U(S,V),  \frac{\partial}{\partial S} U, \frac{\partial}{\partial V} U \right)
	\eeq 
and
	\beq
	\Phi \circ \ell_F \circ \phi^{-1}: (T,V) \mapsto \left(T, V, F(T,V), -\frac{\partial}{\partial T} F, \frac{\partial}{\partial V} F \right),
	\eeq
respectively, defining  different contact polarizations. This is summarized by means of the diagram \eqref{diag.LT}, below

	\beq
	\label{diag.LT}
	\begin{tikzpicture}[]
	\matrix[matrix of math nodes,column sep={65pt,between origins},row
	sep={65pt,between origins},nodes={asymmetrical rectangle}] (s)
	{
	|[name=a1]|T^*\mathcal{U}_U							&			&						&			&|[name=a2]| T^*\mathcal{U}_F							\\
	|[name=b1]|\mathcal{U}_U \subset \mathbb{R}^5				&			& |[name=b2]|(\mathcal{T},\mathcal{D})	&			& |[name=b3]| \mathcal{U}_F \subset \mathbb{R}^5			\\
	|[name=c1]|\psi(\mathcal{E}_U)\subset \mathbb{R}^2	& |[name=c2]|\mathcal{E}_U	&						& |[name=c3]|\mathcal{E}_F	& |[name=c4]|\phi(\mathcal{E}_F)\subset \mathbb{R}^2	\\[-30pt]
	|[name=d1]|C^\infty\left[{\psi(\mathcal{E}_U)}\right]			&			&						&			& |[name=d2]| C^\infty\left[{\phi(\mathcal{E}_F)}\right]			\\
	};
	\draw[->]
			(a2) edge node[above] {\(\left[\phi^S_{\frac{\pi}{2}}\right]^*\)} (a1)
			(a1) edge node [left] {\(\pi_{\mathcal{U}_U}\)} (b1)
			(a2) edge node [right] {\(\pi_{\mathcal{U}_F}\)} (b3)
			(b1) edge[bend left=25] node [above] {\(\phi^S_{\frac{\pi}{2}}\)}(b3)
			(b2) edge node[above] {\(\Psi\)} (b1)
			(b2) edge node[above] {\(\Phi\)} (b3)
			(c2) edge node[above] {\(\ell_U\)} (b2)
			(c3) edge node[above] {\(\ell_F\)} (b2)
			(c2) edge node[above] {\(\psi\)} (c1)
			(c3) edge node[above] {\(\phi\)} (c4)
			(c1) edge node[left] {\(\Psi \circ \ell_U \circ \psi^{-1}\)} (b1)
			(c4) edge node[right] {\(\Phi \circ \ell_F \circ \phi^{-1}\)} (b3)
			(d1) edge node[above] {\( \mathcal{L} \)}(d2)
	;
	\end{tikzpicture}.
	\eeq
Note that, indeed, the relationship between thermodynamic potentials, expressed by the Legendre transformation
	\beq
	\mathcal{L}\left[U \right] = U- T S.
	\eeq
Thus, albeit $[\phi^S_{\frac{\pi}{2}}]$ is merely a change of coordinates for $\mathcal{T}$, it does not represent a diffeomorphism between the corresponding Legendre sub-manifolds. Therefore, a contact polarization is morally tantamount to the choice of a thermodynamic potential, while the various thermodynamic potentials are related by Legendre transformations.

%%%%%%%%%%%%%%%%%%%%%%%%%%%%%%%%%%%%%%%%%%%%%%%%%%%%%%%%%%%%%%%%%%%%
\section*{References}
\bibliographystyle{ieeetr}
\bibliography{AlmostContact_Bib}

\end{document}